# ANALYSIS AND VISUALIZATION OF INDEX WORDS FROM AUDIO TRANSCRIPTS OF INSTRUCTIONAL VIDEOS

*Alexander Haubold and John R. Kender*
Department of Computer Science, Columbia University
500 W. 120th St. 450 CS Bldg. New York, NY 10027, USA
{ahaubold,jrk}@columbia.edu
(212) 939-7146, (212) 939-7115

Abstract

We introduce new techniques for extracting, analyzing, and visualizing textual contents from instructional videos of low production quality. Using Automatic Speech Recognition, approximate transcripts (≈75% Word Error Rate) are obtained from the originally highly compressed videos of university courses, each comprising between 10 to 30 lectures. Text material in the form of books or papers that accompany the course are then used to filter meaningful phrases from the seemingly incoherent transcripts. The resulting index into the transcripts is tied together and visualized in 3 experimental graphs that help in understanding the overall course structure and provide a tool for localizing certain topics for indexing. We specifically discuss a Transcript Index Map, which graphically lays out key phrases for a course, a Textbook Chapter to Transcript Match, and finally a Lecture Transcript Similarity graph, which clusters semantically similar lectures. We test our methods and tools on 7 full courses with 230 hours of video and 273 transcripts. We are able to extract up to 98 unique key terms for a given transcript and up to 347 unique key terms for an entire course. The accuracy of the Textbook Chapter to Transcript Match exceeds 70% on average. The methods used can be applied to genres of video in which there are recurrent thematic words (news, sports, meetings, …)

# 1 Introduction

Summarization and indexing of instructional video is becoming increasingly important with the growing use of recorded audiovisual material in university courses. While some research has focused on lecture browsers using highly controlled visual and textual cues, little attention has been given to analysis of audio transcripts and their structural significance. Presentation slides in the Cornell Lecture Browser [1] are effectively used to build a Table of Contents for a lecture, while Jabberwocky [2] uses them in conjunction with an Automatic Speech Recognizer (ASR) to automatically change slides during a lecture. Other systems, such as the Lecture Explorer [3] and Lecture-on-demand [4] use transcripts for interactive text search queries. Common to all of these systems is their focus on individual lectures.

The analysis of audio data has been investigated with respect to lectures in several instances. The Liberated Learning Project [5] intends to use ASR technology to augment an on-going lecture in real time and provide text transcripts off-line. Some video browsers [6] have already incorporated transcribed data using known techniques, such as TF-IDF. Speech indexing, retrieval, and visualization has enjoyed much attention in domains outside instructional videos, for example SCAN [7] for broadcast news stories.

The goal of this work is to extend a lecture browser's ability to include cross-lecture indexing and referencing, in particular within a full university course with 10 to 30 lectures. We take advantage of the relative ease of comparing textual information across lectures, a characteristic that is more difficult when considering visual data [8]. We first present the methods used in capturing transcripts and discuss the common difficulties encountered in the process. Next, we provide details of the analysis stage and tie in the results with several experimental interactive visualization schemes. We conclude with some future directions, including the incorporation of visual media.

# 2 Data Acquisition

## 2.1 Transcript Generation

For our purposes, we are using course videos from the Columbia Video Network and the commercial Automatic Speech Recognizer IBM ViaVoice to extract transcripts. So far, we have analyzed 7 courses from and related to Computer Science with altogether 183 lectures (230 hours of video); 4 out of these have been analyzed with different instructors' voice trainings for an additional 90 transcripts. Most transcripts contain between 5,000 and 14,000 words with minimal punctuation marks. Depending on the course structure, a semester of videos comprises between 10 and 30 lectures, where each lecture tends to be 70 or 120 minutes long. Video and audio are highly compressed from the originally videotaped classroom environment to fit between 50 Mb and 110 Mb for effective distribution to distance-learning students; this results in uncomfortably poor reproductive quality.

While the lectures are recorded in a controlled environment with several video cameras and a clip-on wireless microphone worn by the instructor, the levels of technological sophistication and invasiveness on teaching style are rather low. This results in a range of audiovisual quality attributes observed in the compressed videos. The microphone, for example, records not only the instructor's voice, but also sounds from writing on the board as well as some ambient noise. The audio quality is furthermore impacted by the instructor's volume level and the position of the microphone with respect to the speaker's mouth. In summary, while the audio track is just passably good enough for human understanding, it proves to be more problematic to an automatic speech recognizer. When applying IBM ViaVoice to the extracted audio track, the Word Error Rate is at approximately 75%. We have computed this value by manually transcribing 2 lectures from different instructors and using them as references.

2.2 Issues of Transcription Accuracy

Glancing over an ASR transcript at first reveals a potpourri of dictionary words, yet a closer comparison to a manual transcript does confirm valid matches of a few (≈25%) distinct phrases. The term "phrase" is used to describe any number of words (≥1) that appear in a semantically meaningful fashion. Table 1 exhibits a section from a typical transcription. Besides a modest portion of correctly recognized words, there exist a large number of unique, yet incorrectly identified words (Nafta, assassinations, …). Using known methods of keyword extraction does not establish the desired separation between correctly and incorrectly recognized text. We will later show how undesirable words can be filtered out by using an external corpus of expected index phrases.

| Manual Transcript (129 words) | Automatic Transcript (103 words) |
|---|---|
| … **deal with** with the data structure like this actually you deal with **it with** with with heaps also so you have some data **structure** right where where items have names **and the question's** how do you how do you get **how** do **you get** to the items we've actually you you should have asked this **question** already **this** semester right uhm so **and** there this data **structure doesn't provide a way to find something** right **like a BINARY TREE provides** if i'm **looking for 27** in a binary tree you know just **given** the **POINTER to the** root of **the** tree I have a way to find it right and if you have an array given you know **the name** of the array you have some way to **find** … | … **deal with** live this *church* is that *CD* in do **it with** wit of the need all sell *Nafta* this **structure** that will write and *assassinations* **and the question is how you** have to **get** at it the added that slate on ye shall ask the **question** *redhead* **this** vast array of Aum sell **and** it is its **structure doesn't provide a way to find something like a BINARY TREE provides** a way of **looking for 27** and by treat it is **given** a **POINTER to the** router **the** *treehouse* where it ought and emulate even though **the name** *Ray Hunt's family* **finds** … |

Table 1: Comparison between manual and automatic transcripts for the course "Analysis of Algorithms". The Word Error Rate is 71%. Matches are highlighted in **bold**. Unique, yet incorrect words are marked with *italics*. Words finally used in the indexing tool are CAPITALIZED.

Other words unknown to the ASR dictionary may be confused with contextually wrong counterparts, e.g. a "lexer" from "lexical analysis" becomes a "laxer". Omission of such words may prove problematic in the already limited collection of accurate text, especially if the phrase is a key term. Additional training and dictionary customization may solve this problem.

Training the software with the instructor's original voice instead of applying some other person's voice for transcribing a lecture resulted in marginal improvements of only 3% for Word Error Rate. At the same time, the raw number of identified index phrases and their occurrence remained approximately the same at < ±1% (see Table 6). However, the qualitative difference between using matched and unmatched voices was more significant. The difference in uniquely identified index phrases from the same lecture was as much as 20%. The benefits of this substantial difference will be discussed later.

While the resulting overall recognition accuracy still remains rather low at 25-30%, we can attribute most of the loss to the poor quality of the recordings. When the 5 instructors who provided training data used the microphone with a Digital Signal Processing unit at a computer, the Speech Recognizer captured most of the spoken words. These results compare to those from the Liberated Learning Project [5]: With intensive voice training and using special microphones and hardware, the transcription accuracy was 80%. How analysis of casual speech and the creation of custom dictionaries from external sources can lead to improvements in speech recognition of lecture material has been investigated in more detail in [9].

Characteristic of lecture speech is its lack of grammatically accurate sentence structure. This includes repetitions (e.g. "how do you how do you get how do you get"), missing sentence completions (e.g. "…get to the items we've <END?> Actually you should have asked this question …"), corrections (e.g. "… so and there This data structure doesn't …"), and filled pauses (uhm, okay, etc.). While this lack of structure in speech does not map to the careful preparation of a material in a textbook, we are still able to use the external corpus of index terms to filter out a small portion of key terms from the transcripts. We will also show how an approximate correspondence can be made between lecture transcripts and chapters from the textbook using word pairs.

## 3 Analysis

### 3.1 Definition of the Target Corpus

For the purpose of indexing, summarization, and cross-referencing, meaningful text needs to be extracted from the transcripts. Ideally, such contents would include "theme" and "topic phrases" that describe the topics covered in a given lecture. We will term them "content phrases". The term "theme phrase" is loosely defined as a phrase shared among several transcripts, i.e. a phrase that appears in at least ¼ of all transcripts, e.g. "data structure". A "topic phrase" denotes the opposite, i.e. a phrase shared in less than ¼ of all transcripts, e.g. "binary tree". The value of ¼ has been experimentally derived from the occurrence patterns displayed in Figure 1. Most index phrases are not repeated in more than ¼ of all transcripts, which makes them good candidates for uniquely descriptive phrases. For example, we could expect the lectures of a course in Computer Science Introduction to Data Structures to have common occurrences of the theme phrases "record", "memory", "insertion", and relatively unique occurrences of the topic phrases "push", "hashtable", "percolate". Theme phrases tend to provide a general tenor for the contents of an entire course or a portion thereof, similar to an abstract of a paper or a back cover summary

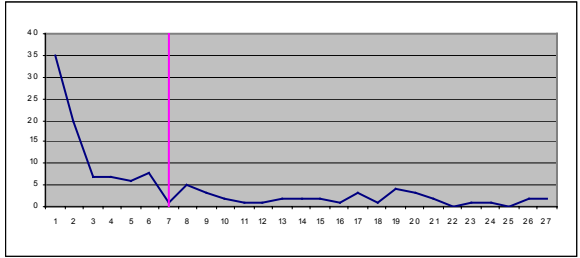
(A) Database Systems

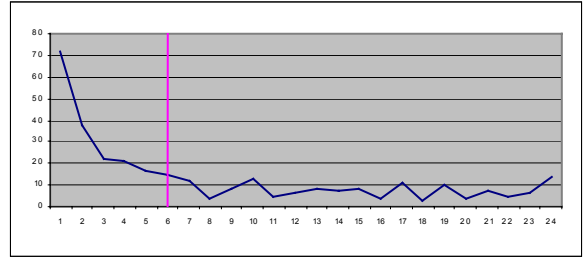
(B) Programming Languages and Translators

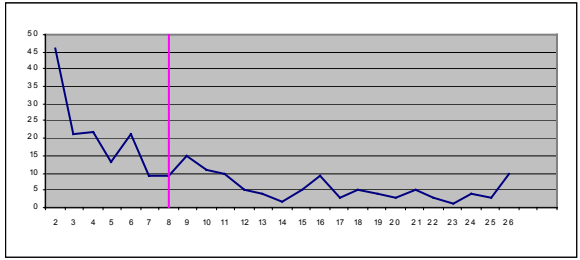
(C) Analysis of Algorithms

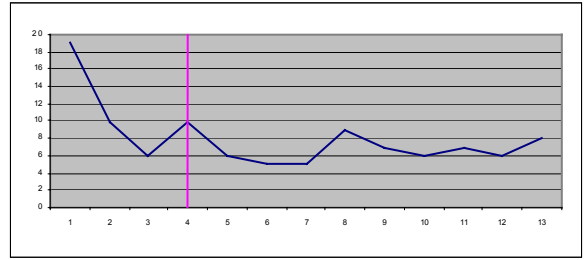
(D) Visual Databases

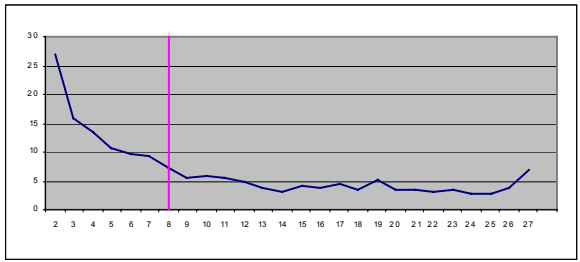
(E) Average of 11 Courses, 273 Transcripts; normalized

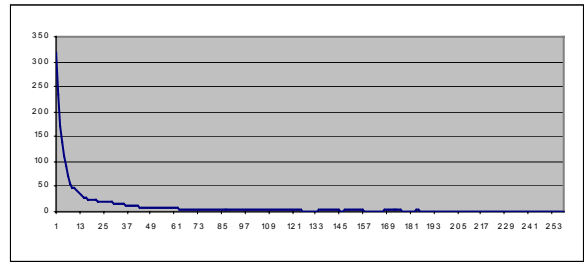
(F) Average of 6 Textbooks, normalized

Figure 1: Index Phrase Dispersion. The y-axes of all graphs denote number of index phrases. The x-axes for (A) through (E) denote number of transcripts, and for (F) number of chapters or sub-chapters. Most index phrases are not repeated in more than ¼ of all transcripts.

of a book. Topic phrases single out specific topics for one or more lectures, as we would expect from a Table of Contents and chapters of a textbook.

A second category of useful terms comprises unique "illustration phrases" used in examples and exercises during class lecture. A topic on scheduling algorithms may, for example, be illustrated by the pipeline in a "car factory", and topics in probability and counting tend to be demonstrated with "red", "green" and "blue marbles". Including these words in a transcript summary and using them to build an index would be highly desirable. Extracting such terms is complicated by three observations. Firstly, illustration phrases tend not to be readily available in a standard external index, which would allow us to efficiently find them. Secondly, the low-accuracy transcripts contain a relatively large amount of wrongly recognized unique words, which cannot easily be distinguished from correctly recognized illustration phrases. Lastly, conversational speech in a classroom environment will necessarily include a fair amount of topic-unrelated anecdotal chat between the instructor, the class, and possibly other parties. The difference between meaningful and meaningless contents cannot be easily discerned without additional cues. We have experimentally applied TF-IDF without significantly successful results; the method captured mostly incorrect terms, as they outweighed the number of correct ones. A possible solution is to ask the instructor to manually add expected illustration phrases to the standard index used for finding content phrases. In our experiments we have added the illustration phrase "make change" to the index of an "Analysis of Algorithms" course, because it was used for specific examples in dynamic programming. Adding the phrases "java" and "gcc" to the index of a Compiler book proved very effective for the final index phrase visualization as well.

## 3.2 Filtering Index Phrases

In order to filter out the larger portion of meaningless text from the ASR transcripts, we obtain a corpus of expected phrases and use it as a dictionary of allowable terms. For the purpose of finding an appropriate corpus for lecture transcripts, we employ the course textbook's index. Since an index generally serves itself as a filter of key phrases for a book, we hypothesize that it can be extended to do the same for lecture transcripts. A large number of phrases found in the index of a textbook are specific enough to fit the curriculum of a course without becoming too generic to fit a lecture in any domain.

The raw index first undergoes some rudimentary word transformations, which will allow for more successful matching to transcripts later on. These transformations are the result of several observations about commonalities between Automatic Speech Recognition, lecture-style speech, and textbook indices. Considerations are made with respect to length of recognized phrases, use of stop words, and grammatical structure. An example of a transformed index is shown in Table 2.

Given the low-accuracy speech recognition of lectures as well as the casual style of speech, the likelihood of capturing a meaningful phrase decreases dramatically with increasing number of words in the phrase (see Table 3). The structure of phrases in a textbook's index tends to reflect this observation: Most index phrases are 1 and 2 words long when disregarding stop words. However, not all lines in an index are self-contained entries. Indentations are commonly used in an index to hierarchically mark sub-expressions which are intended to be concatenated with the parent expression (e.g. Table 2: "amortized analysis" and "accounting method of" become "accounting method of amortized analysis"). Because of the comparatively low probability of finding the 4-word long index phrase instead of two separate 2-word index phrases, the hierarchical index structure is simply discarded. For the purpose of transforming the index into a dictionary for a set of transcripts, every line of the index becomes one phrase.

| amortize analysis | data structure | random number generator |
| account method | aa tree | sort |
| aggregate analysis | augmentation | linear time |
| bob | avl tree | matrix |
| bottom of a stack | binary search tree | problem |

Table 2: Selected index phrases from textbook "Introduction to Algorithms" (Cormen, Leiserson, Rivest, and Stein). Phrases have been stemmed and some stop words have been removed.

| Words in Phrase | Matched Voice (4 Courses, 90 transcripts) | | Unmatched Voice (4 Courses, 90 transcripts) | | Matched & Unmatched Voices (11 Courses, 273 transcripts) | |
| --- | --- | --- | --- | --- | --- | --- |
| 1 | 23741 | 98.15% | 23362 | 98.04% | 59597 | 97.88% |
| 2 | 417 | 1.72% | 432 | 1.81% | 1208 | 1.98% |
| 3 | 30 | 0.12% | 35 | 0.15% | 78 | 0.13% |
| 4 | 0 | 0% | 0 | 0% | 2 | 0.003% |

Table 3: Frequency of Index Phrases with different lengths. Using a matched voice tends to result in slightly more identified index phrases. Unmatched voices, on the other hand, contribute marginally more phrases containing more than 1 word.

The reduction of the index to smaller phrases is also performed with respect to stop words in front and after content words, e.g. "accounting method of" becomes "accounting method", but "call by value" remains the same. Lastly, a Porter stemmer [10, p. 534] is applied to all words. While a full stemmer truncates many words to their absolute and sometimes unintelligible stems, we apply a partial stemmer that only converts plural nouns to singular nouns, and conjugated verbs to their un-conjugated counterparts. Through experimentation, we have observed that a partial stemmer is in fact more effective for this domain of text analysis.

3.3 Filtering Word Pairs

As an alternative to finding index phrases in transcripts, we have explored using word pairs. The rationale behind word pairs is to address the relatively incoherent and fragmented order in which contents occurs within a transcript. Since these fragments are padded with stop words and in many cases with repetitions of stop words, we have defined a word pair as two unordered words appearing anywhere within some fixed distance of another. We have empirically determined this distance to be approximately 10 words for the type of transcripts that we are investigating.

The large number of word pairs (at most ten times the number of words in transcript) that is obtained from this analysis is reduced to a smaller set by filtering each word pair by using the textbook index. Only word pairs where both words appear somewhere in the index are relevant. The resulting list of word pairs is on average one order of magnitude larger than the list of index phrases obtained in (3.2). From the example in Table 4 it is apparent that most word pairs have no coherent semantic meaning, yet some of them do provide some context for the transcript's contents. While they are not useful for visual indexing of transcripts, we find that a correlation can be constructed between their structure and that of a textbook's chapter. One of the user interfaces presented later in this paper discusses how a transcript can be best matched to a chapter in the textbook using word pairs.

Besides using mere occurrence counts of word pairs, we have also employed the $G^2$ log-likelihood statistic to discover significant collocations [11]. As shown in Table 5, the results obtained by using this method are by far more meaningful than word pairs alone. Terms that have not already been filtered by index phrases are added to the final visual index. While the log-likelihood statistic is semantically stronger than simple counting, the latter does perform marginally better in establishing correlations between textbook chapters and transcripts, as discussed later.

| multiple instruction | million low | call structural | clock instruction |
| multiple operation | million improvement | call hazard | clock operation |
| multiple very | million performance | call instruction | clock cpi |
| multiple word | million time | call compaction | clock per |
| multiple processor | million change | call step | clock optimize |

Table 4: Identified word pairs from a "Computer Architecture" course. Some word pairs do not have any semantic meaning, e.g. "multiple very", yet others are easily recognizable, e.g. "clock cpi".

| register file | operand read | register result | number cycle |
| clock cycle | register cycle | order issue | very simple |
| up speed | structure data | cycle instruction | cycle read |
| set block | register instruction | history local | little bit |
| number block | size block | instruction issue | station reservation |

Table 5: Word pairs in decreasing order of log-likelihood. Almost all of these word pairs have an immediately recognizable semantic significance.

3.4 Results for Filtering Index Phrases

In performing our analysis on 273 transcripts, we have been able to identify a reasonable number of index terms in the ASR transcripts (see Table 6 for details). On average, between 30 and 414 index phrases were found for a given transcript, while between 8 and 98 of them were unique occurrences within that transcript. Between 20% and 30% of the index phrases for a transcript had a comparatively significant occurrence between 5 and 50, while between 35% and 50% of them occurred only once. Finally, the number of unique index phrases across an entire course of 10 to 30 lectures was computed to be between 40 and 347 for textbook indices that contained between 253 and 4701 unique index phrases.

While the absolute results with respect to number of index phrases per transcript and unique phrases per course are roughly the same from using two different voice trainings, the qualitative difference is more significant. Table 7 summarizes the improvements for 4 courses; the average number of unique index words per lecture increased up to 18%, while the number of unique words per course saw an increase of up to 10%. The intersection from trained and untrained index phrases turns out to include mostly rare terms that have no useful value in indexing. However, the union of the two eventually is a better set of index phrases to work with.

The low match rate between transcripts and a textbook index of 5% to 11% can be attributed to a number of external factors that cannot be remedied even with perfect transcriptions. Firstly, university courses do not cover all of the material in accompanying textbooks. Specifically, the courses we have surveyed here cover no more than 50% of the reading material. Secondly, indices contain not only content words, but also names of individuals and aliases that most of the time are not mentioned in a lecture. Factoring in these observations and realizing that the transcripts are only 25% accurate, the 5-11% figure is not too unrepresentative.

|  | Course | Avg # Words in Transcripts per Lecture | Avg # Identified Index Phrases per Lecture | Avg # Unique Index Phrases per Lecture | Total # Unique Index Phrases per Course |
|---|---|---|---|---|---|
| Matched Voice | Databases | 6121 | 100 | 33 | 127 |
| | Prog. Lang. | 7446 | 249 | 60 | 202 |
| | Algo. '03 | 7354 | 414 | 98 | 347 |
| | Vis. DB | 13856 | 363 | 50 | 105 |
| Unmatched Voice | Databases | 6182 | 98 | 33 | 130 |
| | Prog. Lang. | 7533 | 258 | 61 | 209 |
| | Algo. '03 | 8061 | 390 | 98 | 336 |
| | Vis. DB | 14013 | 373 | 50 | 102 |
| | Algo. '00 | 8038 | 280 | 70 | 241 |
| | Prob. Stat. | 5927 | 30 | 8 | 40 |
| | Comp. Arch. | 7956 | 159 | 50 | 222 |

Table 6: Statistics for Index Phrase detection averaged over all lectures in a course.

| Course | Avg # Identified Index Phrases per Lecture | % Increase over using only Matched voice | Avg # Unique Index Phrases per Lecture | % Increase over using only Matched voice | Total # Unique Index Phrases per Course | % Increase over using only Matched voice |
|---|---|---|---|---|---|---|
| Databases | 109 | 9% | 39 | 18% | 136 | 7% |
| Prog. Lang. | 260 | 4% | 69 | 15% | 222 | 10% |
| Algo. '03 | 436 | 5% | 116 | 18% | 361 | 4% |
| Vis. DB | 368 | 1% | 53 | 6% | 106 | 1% |

Table 7: Statistics for Index Phrase detection using the combination of results from Matched and Unmatched Voice trainings.

## 4 Results

We have investigated several interactive visualization techniques that present the results from text analysis to the student in a meaningful fashion (see http://www.cs.columbia.edu/~ahaubold/ TranscriptAnalyzer for an interactive demo). The 3 different graphs were developed out of the available dimensions: transcripts, textbook chapters, identified phrases, occurrence of index phrases in transcripts, and occurrence of index phrases in chapters. Because it is up to the student to decide at what level of detail to view the textual contents (theme versus topic), some of the threshold values were incorporated into the user interface as variable sliders.

Common to all 3 visualizations are three parameters that are roughly analogous to a camera's settings. A "zoom" feature allows for setting the specificity of the displayed phrases, ranging from topic-specific to entirely thematic. This measure is derived from the occurrence of a phrase across transcripts, where the zoomed-in topic-specific phrase appears in few transcripts (1 = lowest), and the zoomed-out thematic phrase appears in many transcripts (# transcripts = highest). The "focus" setting denotes the frequency with which a phrase occurs, which is derived from the occurrence of a phrase within a given transcript. The higher the focus is set, the more that outlying and minimally occurring phrases are removed from display. The third common setting, "contrast", controls the length of the phrases considered for display. Increasing this setting bumps out phrases with fewer words, thus creating an emphasis effect on longer phrases.

4.1 Transcript Index Map

The Transcript Index Map is a graph in which index phrases are mapped to the transcripts they appear in. The purpose of this visualization is two-fold. Primarily it is to provide the equivalent of a textbook index to each transcript, except that the index terms are not ordered alphabetically, but rather in order of occurrence. Transcripts appear temporally increasing along the horizontal direction, and index phrases drop vertically below each transcript in decreasing order of occurrence. To further distinguish the frequency with which an index phrase occurs, each item is colored in a spectrum from red to yellow denoting high to low occurrences, respectively. Figure 2 shows an index map in which the zoom value has been set to 1, which effectively displays those terms that appear in no more than 1 transcript. The result is a collection of topic terms per lecture that describe the contents of that lecture as narrowly as possible, e.g. "aggregate analysis", "random number generator", "optimal substructure", etc.

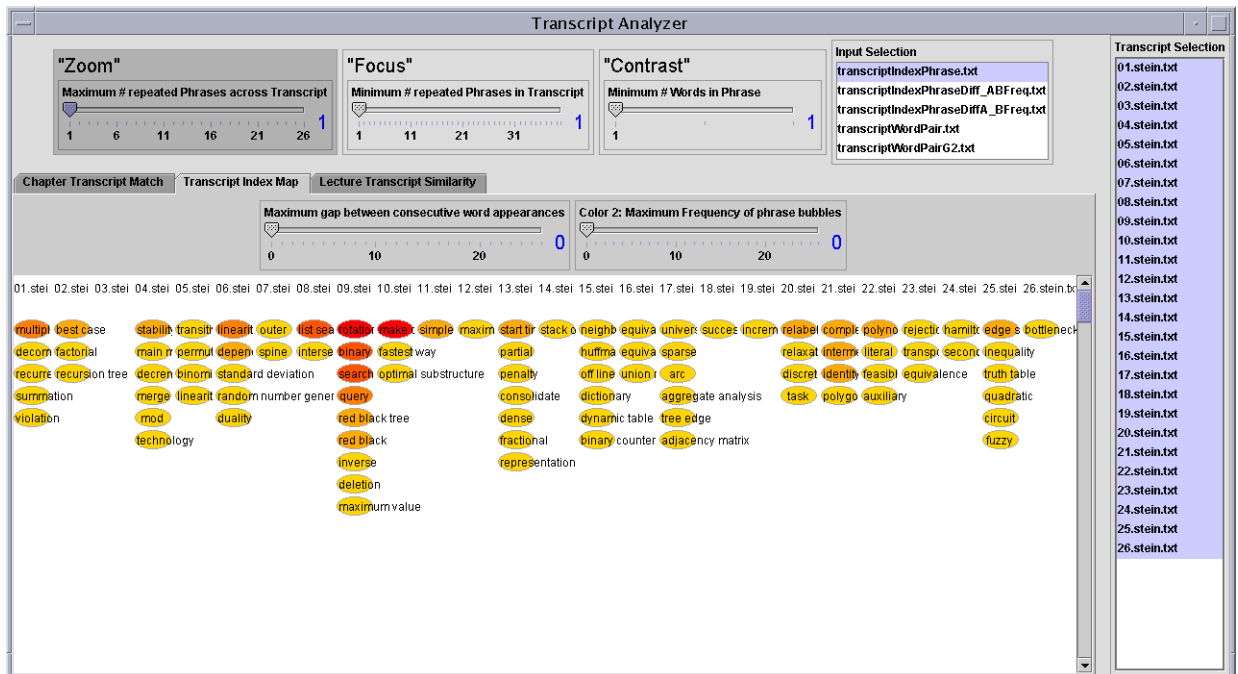

Figure 2: Transcript Index Map for the course "Analysis of Algorithms": Zoom is set to 1, which displays only those topic phrases which occur uniquely in a given transcript.

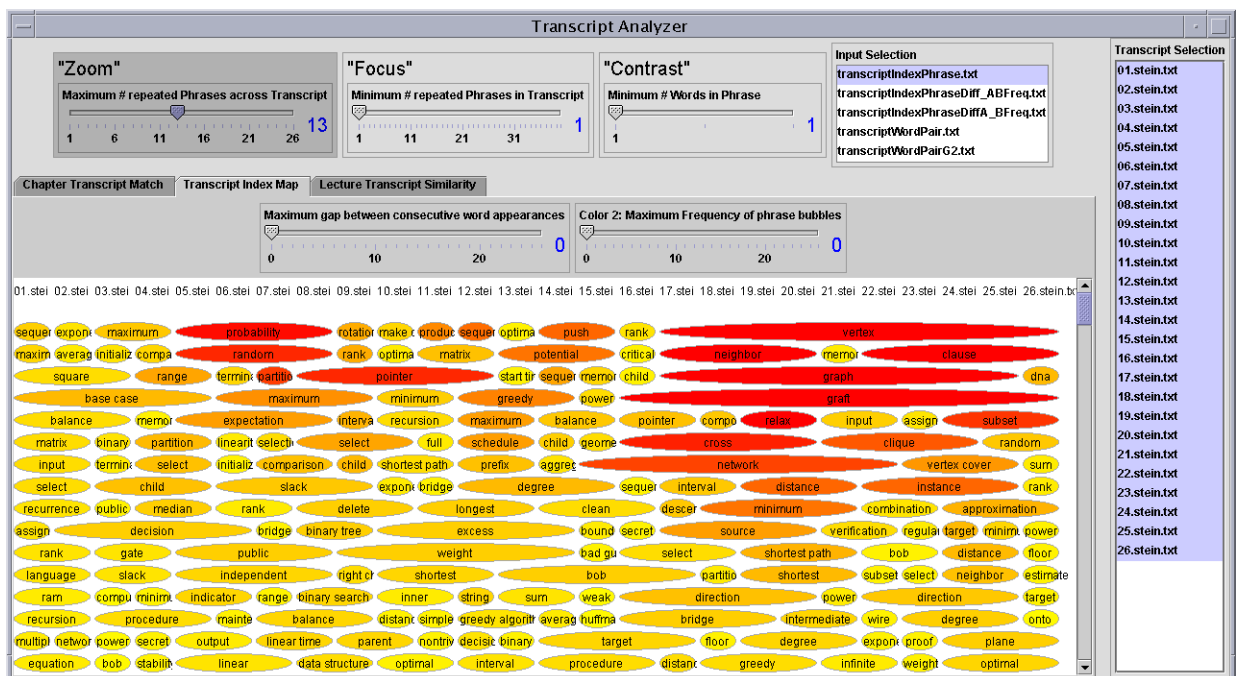

Figure 3: Transcript Index Map: Zoom is set to 13, i.e. half the number of transcripts for this course. Displayed are topic and theme phrases, with theme phrases appearing in larger blobs.

The second function of the Transcript Index Map is to cross-reference index phrases among consecutive transcripts. For this purpose, semantically equal terms are grouped and their occurrence values are summed, effectively increasing their importance in becoming theme phrases. Visually, a grouped item also appears longer, denoting its temporal dependence. An index phrase that appears in 5 consecutive lectures is grouped in one entity that now spans those 5 lectures. As a result, the graph contains differently sized items, which are laid out using a greedy algorithm that fills up as many empty spots as possible nearest to the top. We rationalize this decision by noting that even if the greedy solution is not optimal the relative occurrence of an index phrase is still maintained using color. Figure 3 shows an index map in which the zoom value has been set to 13, which is half the number of available transcripts. Several theme phrases are now readily available: "graph", "vertex", "vertex cover", "shortest path", "probability", etc.

The remaining parameter settings of focus and contrast can be used further narrow down the displayed index. When increasing the value of focus, lower-frequency phrases are removed from the graph, thus "cleaning out" terms that may not be as contextually important due to infrequent use. Increasing the value of contrast removes all phrases with less than a certain number of words. The effect of this setting increases the semantic importance of the displayed phrases, because longer phrases tend to carry more meaning, e.g. "binary search tree" versus "tree".

## 4.2 Textbook Chapter to Transcript Match

In this second visualization we attempt to match a given transcript to a textbook chapter based on the set of identified index phrases. While not every lecture must have a corresponding chapter in the textbook, and while some lectures cover more than one chapter, this interface highlights those chapters that have a relatively high probability of being interesting. Depending on the actual usage of the textbook by the instructor, the display matrix may display a diagonal (see Figure 4) if most chapters in the book are covered in order, or the matrix may display a sparse usage of chapters (see Figure 5).

The tabular interface is divided into individual chapters from the textbook in columns, and lecture transcripts in rows. Each cell represents a numeric value that ranks the relative score for each chapter-transcript pairing. The score is based on a conceptual three dimensional histogram, whose first dimension is transcript number, second dimension is chapter, and third dimension is phrase identifier (varying from 1 to total number of phrases in the course). This histogram reorders for $phrase_k$ the number of times it simultaneously occurs in $transcript_i$ and $chapter_j$, each histogram bin thus is named count(i, j, k). We define

$$score(i, j) = \sum_k \ln(count(i, j, k))$$

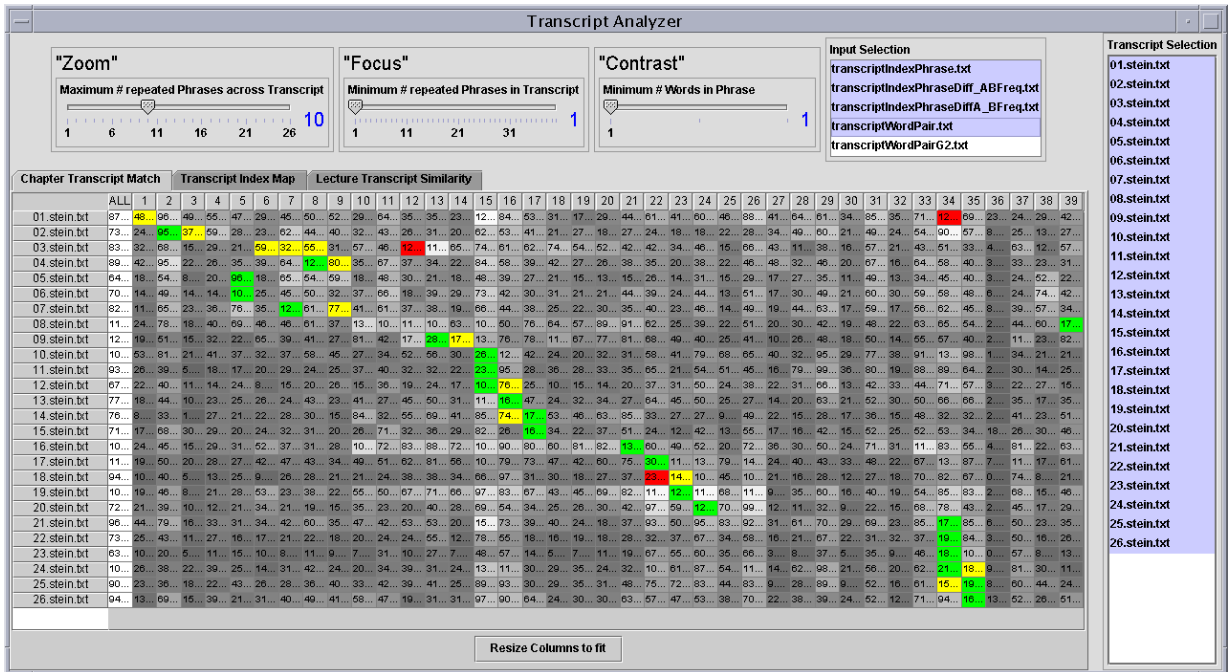

Figure 4: Chapter Transcript Match for the Course "Analysis of Algorithms": The instructor follows the book in order, which can be seen from the diagonal. The outlier in the rightmost column is additional reading material that was not covered in the book. Green cells denote correct matches of transcripts to chapters. Yellow cells denote other valid correspondences, although only the most likely one is chosen by the interface. Red cells denote incorrect matches.

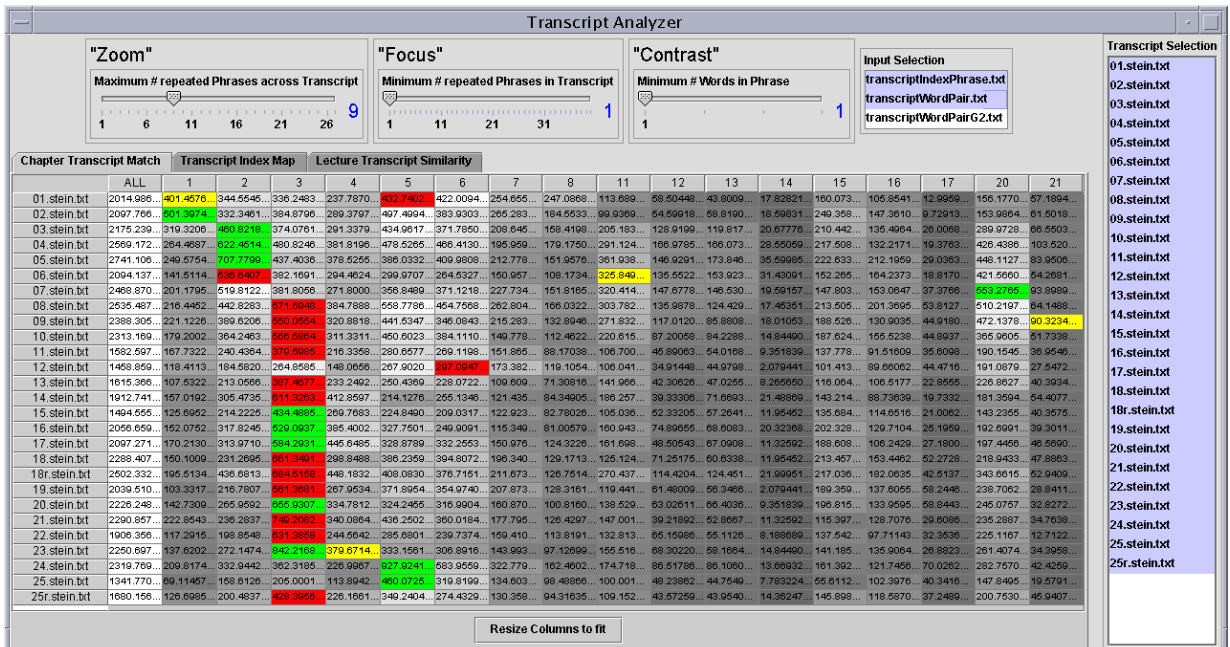

Figure 5: Chapter Transcript Match for the Course "Computer Architecture": The instructor focuses mostly Chapters 2 and 3 of the book and some additional reading material (2 right-most columns).

That is: For every phrase in a given transcript i, add the logs of the occurrences of that phrase in chapter j; this approximates a joint probability measure.

We studied alternative ways of computing the transcript-chapter match: Instead of using counts of simple phrases, we looked at three different word sets. We investigated index phrases, word pairs, and word pairs that had a high $G^2$ score (i.e. collocations). Figure 6 summarizes the qualitative difference among these 3 sets, for 5 courses with altogether 107 lectures, 93 of whose ground truth assignment to one or more chapters in the textbook was obvious. Using index phrases alone, about 50% of the lectures could be matched to the correct chapter using a zoom value between 6 and 17. Word pairs by themselves achieved around 66% of correct matching in a zoom range between 14 and 26. Using word pairs derived from the $G^2$ measure performed marginally worse at 63%. The combination of index phrases and word pairs resulted in the best average matching rate of 70%. Remarkable is also the robustness at different zoom levels. The range of matching results when disregarding the extreme start and end points is between 61% and 78%.

While textbooks have clear definitions of chapters and sub-chapters, it is unclear what exactly constitutes a "chapter" with respect to this visualization tool. In Figure 4 there exists a clear sense of correspondence between chapters and transcripts, while in Figure 5, several lectures span one chapter. Individual columns could be split into sub-chapters; however, we found that the accuracy of matching drops about 50%, namely due to the sparsity of book text (1 chapter ≈ 10 sub-chapters).

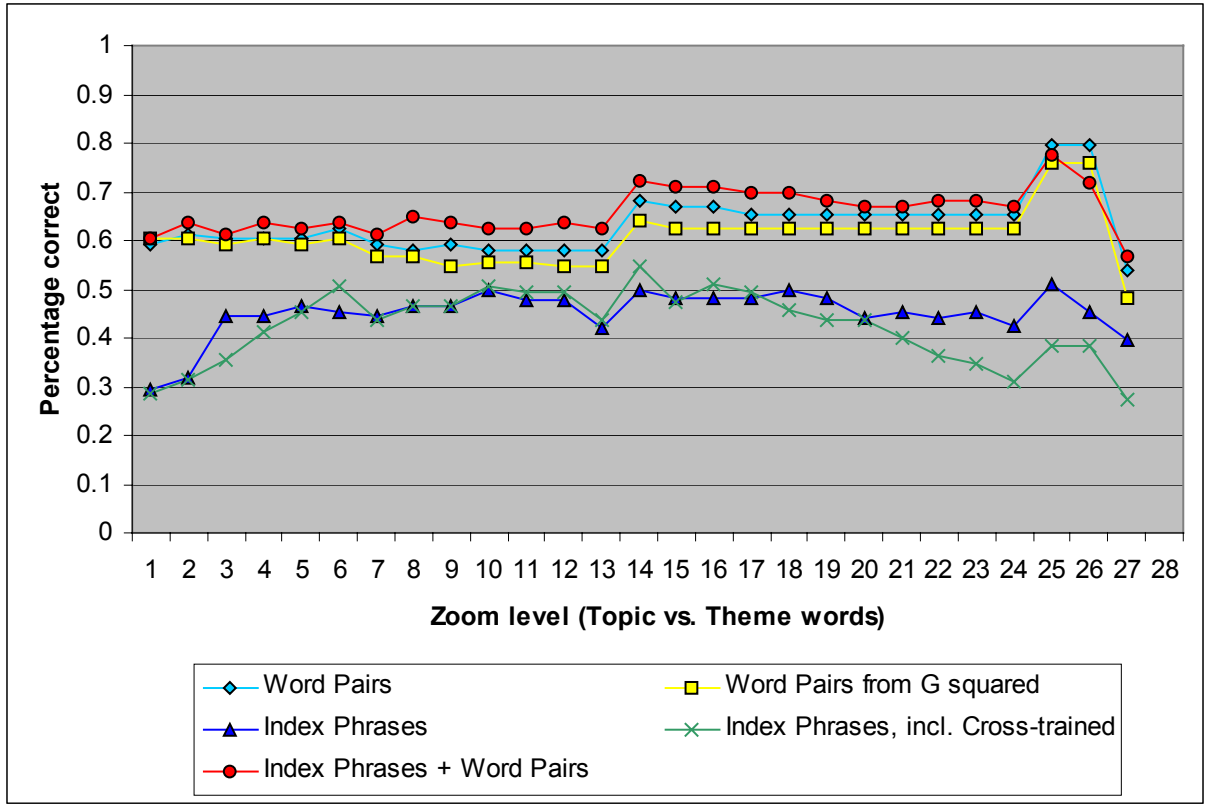

Figure 6: Chapter Transcript Matching: Word Pairs, and the combination of Index Phrases and Word Pairs, perform best in matching chapters to transcripts.

4.3 Lecture Transcript Similarity

For the third visualization of lecture contents for a full course, we have created a graph that visually clusters similar lectures based on a set of selected phrases. The purpose of this tool is to allow a student to explore a course by dynamically grouping lectures that have similar contents based only on a small set of index phrases (see Figure 7). Closely related transcripts are clustered and linked in red. Weakly related transcripts are linked with a color that fades into the background, while unrelated transcripts are not linked at all.

Multidimensional Scaling is used to collapse the higher dimensional space of N lecture transcripts down to 2 dimensions. The distance matrix used for MDS is constructed by means of the Dice Distance applied to each pair ($i,j$) of all transcripts:

$$dist(i, j) = \frac{b+c}{2a+b+c}$$

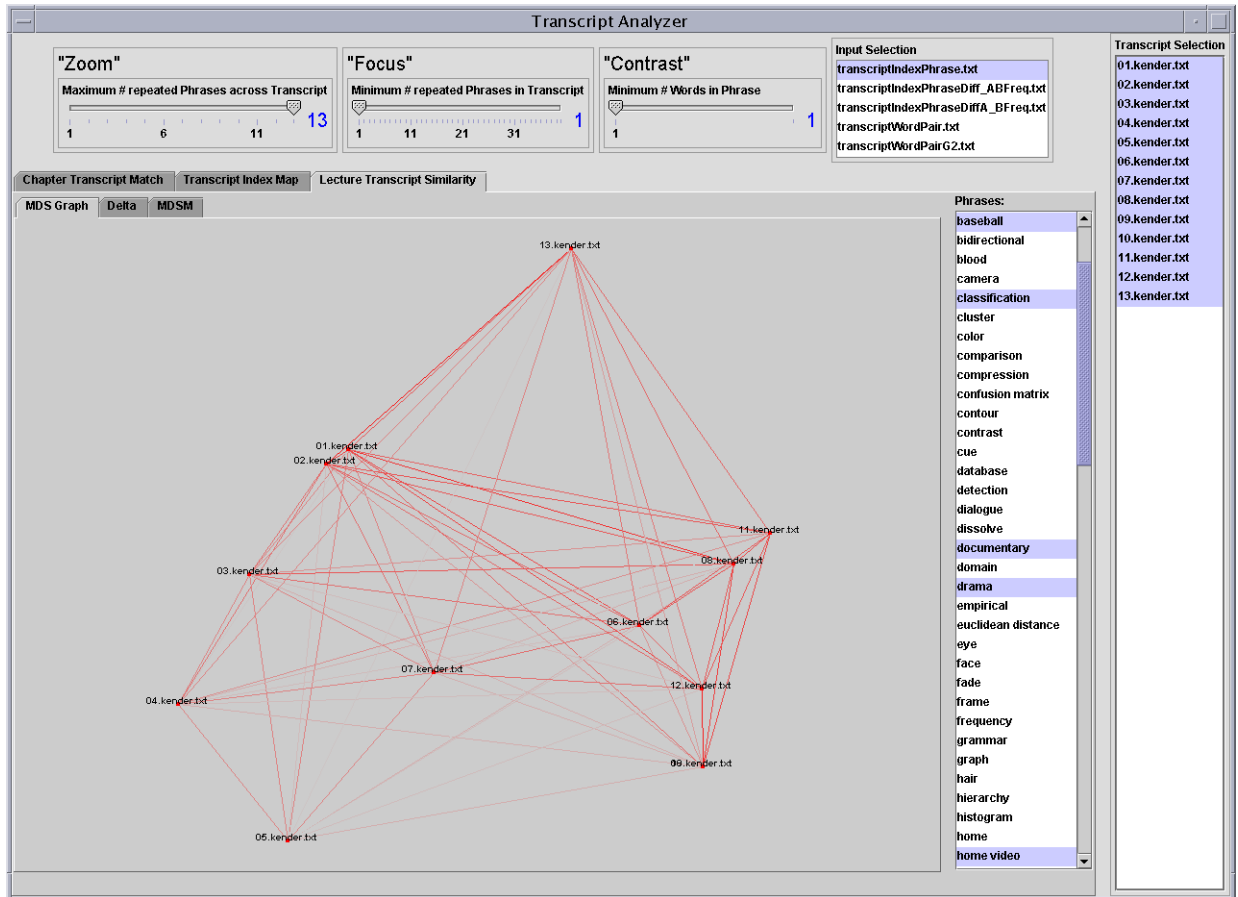

Figure 7: Multidimensional Scaling of transcript similarity based on a selection of index phrases. Lectures on video classification (baseball, documentary, drama, etc.) are clustered near the right, while lectures related to image analysis are closer to the left. In-between is a mixed lecture on both topics. The outlier close to the top is a review session for the entire course.

where a, b, and c are the co-occurrence counts of all phrases (a) in transcript i and j, (b) not in transcript i but in transcript j, and (c) in transcript j but not in transcript i.

We have found that semantically meaningful contents, such as index phrases, produce distinguishable graphs. Closely related lectures appear in clusters, while largely unrelated lectures produce outlier nodes. Figure 7 shows a graph for the selection of phrases "baseball", "classification", "documentary", "drama", "home video", "musical", "newscast", "sitcom", "soccer", and "video" from a course in "Visual Databases", which covers topics on image and video analysis, retrieval, summarization, and visualization. These video classification terms appear mostly in lectures 6, 8, 9, 10, 11, and 12, which can be seen clustered on the right of the graph (Note: 9 and 10 overlap). Lectures 3, 4, and 5 cover image retrieval and face recognition and thus appear farthest away near the left of the graph. Centered between these two clusters we find lecture 7, which discusses jpeg and mpeg algorithms; this also corresponds to a "semantic average" between images and video. An outlier in this visualization is lecture 13 near the top; it serves as a review session of the entire course.

## 5 Conclusion and Future Directions

We have presented new methods for extracting meaningful textual information from low-accuracy lecture transcripts using an external corpus of index phrases. Interactive visualizations show that these methods can be a very useful addition to course lecture browsers. More importantly, our analysis of transcripts shows how the easily obtained data can be employed to provide a higher-level structure of an entire course made up of several (10 to 30) lectures, as opposed to restricting the data to individual lectures.

In the near future, we will be conducting user studies on the interfaces, after incorporating the tools into our previously developed lecture browser based on the visual structure of the videos [8], pictured in Figure 8. Additional interfaces are being explored for visualizing the textual information on a finer grained time scale. The inclusion of lecture notes, presentation slides, and other course materials may benefit the already good results of Chapter Transcript Matching. We also plan to test our methods on courses from departments unrelated to Computer Science.

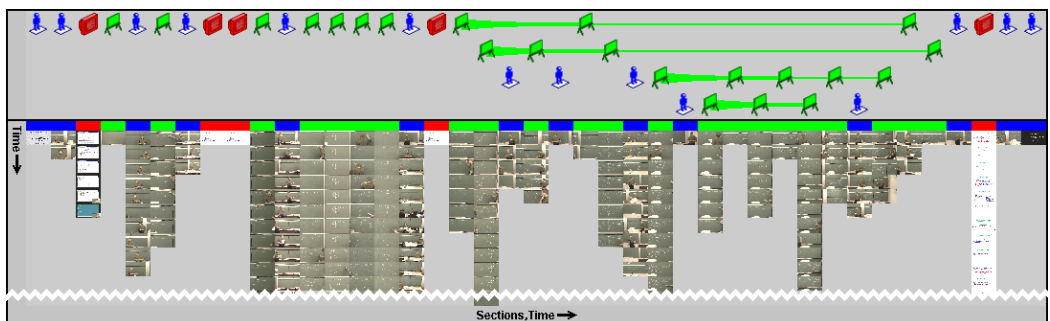

Figure 8: Visual Lecture Browser which will be augmented with text-indexing tools.